\documentclass[11pt]{article}
\usepackage{geometry}

\usepackage{amsfonts}
\usepackage{mathrsfs}
\usepackage{graphicx}
\usepackage{amsmath,amsthm,amssymb,amscd}
\usepackage[all]{xy}
\usepackage{hyperref}
\usepackage{multirow}
\usepackage{color}
\usepackage{float}
\usepackage{mathtools,slashed}
\usepackage[latin1]{inputenc}
\usepackage{dsfont}
\usepackage{empheq}
\usepackage{multirow}
\usepackage{bbm}

\newcommand{\nn}{\nonumber}
\newcommand\fft[2]{\frac{#1}{#2}}
\newcommand\ft[2]{{\textstyle\frac{#1}{#2}}}

\def\eqa{\begin{eqnarray}}
\def\eqae{\end{eqnarray}}
\def\eq{\begin{equation}}
\def\eqe{\end{equation}}
\def\be{\begin{equation}}
\def\ee{\end{equation}}
\def\bea{\begin{eqnarray}}
\def\eea{\end{eqnarray}}
\def\ba{\begin{array}}
\def\ea{\end{array}}
\def\bd{\begin{displaymath}}
\def\ed{\end{displaymath}}

\def\>{\rangle}
\def\<{\langle}

\numberwithin{equation}{section}

\DeclareMathOperator{\Li}{Li}

\begin{document}

\begin{titlepage}
\hfill MCTP-16-19
\begin{center}

{\Large \textbf{Higher Rank ABJM Wilson Loops from Matrix Models}}\\[4em]

\renewcommand{\thefootnote}{\fnsymbol{footnote}}

{\large Jonathan Cookmeyer${}^{a}$, James T. Liu${}^b$
and Leopoldo A.~Pando Zayas${}^{c}$}\\[3em]

\renewcommand{\thefootnote}{\arabic{footnote}}
${}^a$\emph{Haverford College\\ 370 Lancaster Avenue, Haverford, PA 19041, USA}\\[1em]

${}^{b,c}$\emph{Michigan Center for Theoretical Physics,  Randall Laboratory of Physics\\ The University of
Michigan,  Ann Arbor, MI 48109, USA}\\[1em]

${}^c$\emph{The Abdus Salam International Centre for Theoretical Physics\\ Strada Costiera 11,  34014 Trieste, Italy}\\[6em]

\abstract{}
We compute the vacuum expectation values of $1/6$ supersymmetric Wilson loops in higher dimensional representations of the gauge group in ABJM theory. We present results for the $m$-symmetric and $m$-antisymmetric representations by exploiting standard matrix model techniques. At leading order, in the saddle point approximation, our expressions  reproduce holographic results from both D6 and D2 branes corresponding to the antisymmetric and symmetric representations, respectively. We also compute $1/N$ corrections to the leading saddle point results. 
\end{center}

\end{titlepage}


\section{Introduction}

Wilson loops are gauge-invariant non-local operators that can be defined in any gauge theory. These operators provide a window into the dynamics of the theory and serve as important order parameters. For example, in confining theories their expectation values display an area law behavior implying a linear quark-antiquark potential.  In non-confining theories, such as ${\cal N}=4$ supersymmetric Yang-Mills (SYM), the expectation value of Wilson loops has been shown to correspond to a Coulomb interaction. Even in this simpler case much can be learned from the nontrivial dependence on the coupling  constant. 

In the context of the AdS/CFT correspondence, Wilson loops play a particularly important role as they are described, at leading order, by classical configurations of strings and branes \cite{Maldacena:1998im,Rey:1998ik}. These classical configurations represent a controlled departure from the strict supergravity limit into stringy aspects of the correspondence. Indeed, the AdS/CFT dictionary has been enlarged to include D3 and D5 branes  corresponding to Wilson loops in the symmetric and antisymmetric representations of $SU(N)$ for ${\cal N}=4$ SYM  \cite{Drukker:2005kx,Gomis:2006sb,Gomis:2006im,Yamaguchi:2006tq,Hartnoll:2006is}.  

The AdS/CFT correspondence conjectures a mathematical equivalence between string theories and gauge theories. One of the prototypical pairs is string theory on  $\textrm{AdS}_4\times {\mathbb CP}^3$, with Ramond-Ramond fluxes and an ${\cal N}=6$ Chern-Simons theory coupled to matter known as ABJM   \cite{Aharony:2008ug}. Using supersymmetric localization techniques, it was shown in \cite{Kapustin:2009kz} that the computation of the expectation values of some supersymmetric observables in ABJM theory can be reduced to a matrix integral. One of the first observables tackled with this approach was precisely a $1/6$ supersymmetric Wilson loop.  More general results, including exact expressions in the rank of the gauge group, $N$, and the Chern-Simons level, $k$, for other supersymmetric Wilson loops   were obtained using advanced matrix model techniques \cite{Marino:2009jd,Klemm:2012ii}.  

Given their prominent role in the case of the correspondence between strings on $\textrm{AdS}_5\times S^5$ and ${\cal N}=4$ SYM, it is natural to turn our attention to Wilson loops in higher dimensional representations for the case of the correspondence between string theory on $\textrm{AdS}_4\times \mathbb{CP}^3$ and ABJM theory. 
For the most part, the vacuum expectation values of Wilson loops in high-rank representations have not been systematically studied, although some results were reported in, for example,  \cite{Hatsuda:2013yua,Hatsuda:2016rmv}. In this manuscript we use standard matrix model techniques to compute  the leading order expression for the Wilson loops in the $m$-symmetric and $m$-antisymmetric representations in the large-$N$ limit with $f\equiv m/N$ fixed.  We find precise agreement with the holographic results. Namely, our matrix model results match the classical actions of the corresponding D6 and D2 branes in $\textrm{AdS}_4\times \mathbb{CP}^3$  as computed in \cite{Drukker:2008zx}. We also go beyond the saddle point approximation and compute some sub-leading corrections, setting the stage for potential precision tests on the holographic side.

The rest of the paper is organized as follows. In section~\ref{Sec:Prelim} we review the $1/6$ supersymmetric Wilson loop in ABJM theory and describe the general computational setup.   In section~\ref{Sec:Am} we derive the result for  the antisymmetric representation.  We present the details of the symmetric representation in section \ref{Sec:Sm}.   We conclude in section~\ref{Sec:Conclusions}.  We reserve  appendix~\ref{App:Numerical} for a few intuition-building numerical vignettes related to the various approximations used in the main text.

\section{Wilson loops in ABJM theory}\label{Sec:Prelim}%
The ABJM theory is a three-dimensional Chern-Simons-matter theory with $U(N)\times U(N)$ gauge group \cite{Aharony:2008ug}. The gauge fields are governed by  Chern-Simons actions
with opposite integer levels for the two gauge groups, $k$ and $-k$. The matter sector contains four complex scalar fields $C_I, (I=1,2,3,4)$ in the bifundamental representation $({\bf N}, \bar{\bf N})$ and the corresponding complex conjugate in the $(\bar{\bf N}, {\bf N})$ representation; the theory also contains fermionic superpartners  (see \cite{Aharony:2008ug}  for details). 

To build $1/6$ supersymmetric  Wilson loops,  one considers only one of the gauge fields of the whole $U(N)\times U(N)$ gauge group, denoted by   $A_\mu$. To preserve supersymmetry we need to include a contribution from the matter sector. The main intuition comes from the construction of  supersymmetric Wilson loops in ${\cal N}=4$ SYM. However, in the absence of adjoint fields, one considers the appropriate combination of bi-fundamentals, $C_I$, namely \cite{Drukker:2008zx,Rey:2008bh,Chen:2008bp}:
\be
W_{\cal R}=\frac{1}{{\rm dim}[{\cal R}]}{\rm Tr}_{\cal R}\, {\cal P}\int \left(i A_\mu \dot{x}^\mu +\frac{2\pi}{k} |\dot{x}|\, M^I_J C_I \bar{C}^J\right) ds,
\ee
where ${\cal R}$ denotes the representation. It was shown in \cite{Drukker:2008zx,Rey:2008bh,Chen:2008bp} that the above operator preserves $1/6$ of the 24 supercharges when the loop is a straight line or a circle and the matrix takes the form $M^I_J={\rm diag}\, (1,1,-1,-1)$.

A remarkable result of \cite{Kapustin:2009kz} was to show that the computation of the vacuum expectation values of these Wilson loops reduces to a matrix model. Namely,  for supersymmetric observables, it suffices to compute their expectation using the following partition function:
\begin{equation}
Z(N,k) = \frac{1}{(N!)^2} \int\! \prod_{i=1}^N \frac{d\mu_i}{2\pi}\frac{d\nu_i}{2\pi}
\frac{\prod_{i<j}\left(2\sinh\frac{\mu_i-\mu_j}2\right)^2\left(2\sinh\frac{\nu_i-\nu_j}2\right)^2}{\prod_{i,j}\left(2\cosh\frac{\mu_i-\nu_j}2\right)^2}
\exp\!\left[\frac{ik}{4\pi}\sum_i (\mu_i^2-\nu_i^2)\right]\!.
\label{eq:ABJMZ}
\end{equation}

The Wilson loop in the symmetric, $S_m$, and antisymmetric, $A_m$, representations are given by the following expression in terms of the eigenvalues $\mu_i$  (for the other gauge group   the eigenvalues $\nu_i$ would be involved):
\begin{equation}
W_{1/6}^{S_m} = \frac{1}{\dim[S_m]}\sum_{1\le i_1\le \cdots\le i_m\le N} \exp[\mu_{i_1}+\cdots+\mu_{i_m}], \end{equation}
\begin{equation}
W_{1/6}^{A_m} = \frac{1}{\dim[A_m]}\sum_{1\le i_1< \cdots< i_m\le N}  \exp[\mu_{i_1}+\cdots+\mu_{i_m}],
\end{equation}
where $\dim[S_m]$ and $\dim[A_m]$ are the dimensions of the respective representations. Note that the main difference is in the ordering of the eigenvalues. A convenient way of accessing these
operators, as noted in \cite{Hartnoll:2006is}, is to use the generating functions for the symmetric and antisymmetric representations, respectively
\begin{equation}
F_S(t) \equiv \prod_{i=1}^N \frac{1}{1-te^{\mu_i}}, \qquad \text{ and}\qquad
F_A(t) \equiv \prod_{i=1}^N (t+e^{\mu_i}).
\end{equation}
The expectation values of the Wilson loops are then the coefficients of the appropriate powers of $t$.  One efficient way of extracting the vacuum expectation value of Wilson loops from the generating functions is by performing contour integrals
\begin{equation}
W_{1/6}^{S_m} = \frac{1}{\dim[S_m]}\frac{1}{2\pi i}\oint_{\mathcal C_0} \frac{F_S(t)}{t^{m+1}}dt ;
\qquad W_{1/6}^{A_m} =\frac{1}{\dim[A_m]}\frac{1}{2\pi i}\oint_{\mathcal C_\infty} \frac{F_A(t)}{t^{N-m+1}}dt,
\end{equation}
where $\mathcal C_0$ is around zero and $\mathcal C_\infty$ is around infinity.

We will focus on the planar limit in the large-$N$ expansion.  In preparation for the limit and with a view toward using the steepest descent method, we write  $\langle F_{S,A}(t)\rangle$ as:

\begin{equation}
\langle F_{S,A}(t)\rangle=\frac{1}{Z}
\int \prod_i \frac{d\mu_i}{2\pi} \frac{d\nu_i}{2\pi} \exp(-S_{A,S}),
\label{fsa}
\end{equation}
with
\begin{align}
S_{S,A} = -\frac{ik}{4\pi}\sum_i (\mu_i^2-\nu_i^2) - &\sum_{i<j} 2 \ln\left[ \left(2\sinh\frac{\mu_i-\mu_j}2\right)\left(2\sinh\frac{\nu_i-\nu_j}2\right)\right]\\
+&\sum_{i,j}2\ln\left[2\cosh\frac{\mu_i-\nu_j}2\right]-\sum_i 
\begin{cases} 
- \ln (1-t e^{\mu_i})\\ \ln (t+e^{\mu_i})
\end{cases}.
\end{align}
The top case in the final sum is for symmetric and the bottom for antisymmetric. The steepest descent
method corresponds to evaluating the above integral at the saddle point, given by the equations
$\partial S_{S,A}/\partial\mu_i=\partial S_{S,A}/\partial\nu_i=0$.  Although we have inserted the generating
function $F_{S,A}$ into the integral, we expect that the saddle point solution will be unchanged from
that of the partition function, as the dominant terms have $\mathcal O(N^2)$ dependence, while the
added term, coming from the Wilson loop, has only  $\mathcal O(N)$ dependence.
More rigorously, because the operators $F_{S,A}$ have no $\nu_j$ dependence, only
the $\mu_i$ equations are changed, and become
\begin{equation}
0=\frac{\partial S_{S,A}}{\partial \mu_i} = -\frac{ik}{2\pi} \mu_i - \sum_{j\ne i}\coth\frac{\mu_i-\mu_j}2 +\sum_j \tanh\frac{\mu_i - \nu_j}{2} -
\begin{cases}
\frac{t e^{\mu_i}}{1-te^{\mu_i}} \\
\frac{e^{\mu_i}}{t+e^{\mu_i}}
\end{cases}.
\end{equation}
In principle the parameter $t$ is a formal expansion parameter. However, note that, for small enough $t$ in the symmetric case and for large enough
 $t$ in the antisymmetric case, the added
term is small. Since we are only interested in contours around zero and infinity for the symmetric and
antisymmetric case respectively, it is enough for the eigenvalue solution to remain the same for
small $t$ and large $t$, respectively. Once we make this approximation, we do not need $t$ to be small
or large as we can treat it as a normal contour  integral. Note that at this point we have not yet made any assumptions about the Chern-Simons level,  $k$,  or the 't Hooft coupling, $\lambda=N/k$.  Thus the
saddle point expressions are applicable to both the Type IIA limit  ($N,k\to\infty$ and $\lambda=N/k$ fixed) and the M-theory limit ($N\to \infty$ and $k$ fixed).

\subsection{Eigenvalue distribution}
One key attribute of large-$N$ methods in matrix models is the assumption that  in the limit of large $N$, one can describe the discrete set of eigenvalues by a continuous distribution.  It is, indeed, this distribution of eigenvalues that plays the central role. In this subsection we discuss in detail the approximation we use for the eigenvalue distribution. 

The eigenvalue distribution for the ABJM matrix model in the planar limit was worked out in
\cite{Marino:2009jd,Aganagic:2002wv,Halmagyi:2003ze,Halmagyi:2003mm}, and it  is given by
\begin{equation}
\rho(\mu) = \frac{1}{\pi t_1} \tan^{-1}\left[\sqrt{\frac{\alpha - 2 \cosh \mu}{\beta + 2 \cosh \mu}}\right] \qquad \mu \in[-\mu_*,\mu_*],
\label{eq:rhomu}
\end{equation}
where 
\begin{equation}
\mu_*= \ln\left[
\frac12\left(\alpha+ \sqrt{\alpha^2-4}\right)\right].
\label{Avk}
\end{equation}
While this distribution was derived in the corresponding lens space matrix model, it can be analytically
continued to the ABJM slice by taking
\begin{equation}
t_1 = 2\pi i \lambda, \qquad
\alpha = 2 + i \kappa, \qquad \beta = 2 - i \kappa,
\end{equation}
and 
\begin{equation}
\lambda = \frac{\kappa}{8\pi} \;{}_3F_2\left(\frac12,\frac12,\frac12;1,\frac32;-\frac{\kappa^2}{16}\right),
\label{eq:hyperPFQ}
\end{equation}
where ${}_3F_2$ is a generalized hypergeometric function.  (See Appendix~\ref{App:Numerical}
for additional comments on the analytical continuation and comparison with numerical results.)

Although (\ref{eq:rhomu}) is valid for arbitrary values of $\lambda$, we focus on the Type IIA limit with
large $\lambda$.  In this case, the expression for $\rho(\mu)$ simplifies to be approximately constant,
which is the same as in the M-theory limit.  In particular, for $\lambda\gg1$, we may invert
(\ref{eq:hyperPFQ}) to obtain
\begin{equation}
\kappa = e^{\pi \sqrt{2\hat\lambda}}\left[1+O\left(e^{-2\pi \sqrt{2\hat\lambda}}\right)\right]\gg1,
\label{kvl}
\end{equation}
where $\hat\lambda\equiv\lambda-1/24$.
The $\mu$-cut then extends from $-\mu_*$ to $\mu_*$ where
\begin{equation}
\mu_* =\pi \sqrt{2\hat\lambda} + i\frac{\pi}{2} + O\left(e^{-\pi\sqrt{2\hat\lambda}}\right).
\end{equation}
The eigenvalue density (\ref{eq:rhomu}) can now be expanded for $\kappa\gg1$, with the result
\begin{equation}
\rho(\mu)\approx\fft1{4\pi^2\lambda}\ln\left(e^\mu+e^{-\mu}-e^{\mu_*}\right),
\label{eq:rhoapprox}
\end{equation}
up to $\mathcal O(1/\kappa)$ corrections, at least for $\mu$ not within $\mathcal O(1/\kappa)$ of the branch point of the log in (\ref{eq:rhoapprox}).  Note that $\rho(\mu)$ remains normalized up to exponentially small corrections in $\lambda$
\begin{align}
\int_{-\mu_*}^{\mu_*}\rho(\mu)d\mu\approx2\int_0^{\mu_*}\fft1{4\pi^2\lambda}\ln(e^\mu-e^{\mu_*})d\mu
&=\fft{(\mu_*-i\pi/2)^2+\pi^2/12}{2\pi^2\lambda}+\mathcal O(e^{-\mu_*})\nn\\
&=\fft{\hat\lambda+1/24}\lambda+\mathcal O(e^{-\pi\sqrt{2\hat\lambda}}).
\end{align}

We further note that, for $\mu$ along the line connecting $-\mu_*$ to $\mu_*$ (but not within $\mathcal O(1/\kappa)$ of the endpoints), the $e^{\mu_*}$ term dominates, and $\rho(\mu)$ may be approximated by the constant distribution
\begin{equation}
\rho(\mu)\approx\fft{\mu^*-i\pi+\mathcal O(1/\ln\kappa)}{4\pi^2\lambda}=\fft{\pi\sqrt{2\hat\lambda}-i\pi/2+\mathcal O(1/\sqrt{\hat\lambda})}{4\pi^2\lambda}.
\end{equation}
Therefore, at leading order in large $\lambda$, in the IIA limit, we recover a constant eigenvalue density
along the line stretching from $-\mu_*$ to $\mu_*$:
\begin{equation}
\rho(\mu)=\fft1{2\mu_*},\qquad\mu\in[-\mu_*,\mu_*].
\label{rhomuc}
\end{equation}
At this order, we do not make a distinction between $\lambda$ and $\hat\lambda$, so we have simply
$\mu_*=\pi\sqrt{2\lambda}+i\pi/2$.  This distribution is the same to leading order in $\lambda$ as the one
in the M-theory limit derived in \cite{Herzog:2010hf}. We therefore expect our results to be applicable in both the large-$\lambda$ IIA limit and the M-theory limit.

\subsection{Large $N$ saddle point approximation}

In the planar limit, the expectation values of the generating functions \eqref{fsa} are simply
\begin{equation}
\langle F_A \rangle \approx \exp\left[ N \int_{-\mu_*}^{\mu_*}d\mu \,\rho(\mu) \ln (t+e^{\mu})\right],
\qquad \langle F_S \rangle \approx \exp\left[- N \int_{-\mu_*}^{\mu_*}d\mu \,\rho(\mu)\ln (1-te^{\mu})\right],
\label{eq:FAFS}
\end{equation}
where the eigenvalue density is given by \eqref{eq:rhomu}, or in the large-$\lambda$ limit by \eqref{rhomuc}.
For the antisymmetric integral, we can do a change of variables sending $t\to 1/t$ just as in \cite{Hartnoll:2006is}, so we can write succinctly (up to a minus sign in the antisymmetric case)
\begin{equation}
\langle W_{1/6}^{S_m,A_m}\rangle = 
\frac{1}{\dim[\mathcal R]}\frac{1}{2\pi i}\oint_{\mathcal C} dt \frac{1}{t^{m+1}}\exp
\left[\mp N \int_{-\mu_*}^{\mu_*} d\mu\, \rho(\mu) \ln (1\mp t e^{\mu})\right]\equiv \frac{I^{S_m,A_m}}{\dim[\mathcal R]},
\label{Wtint}
\end{equation}
where the contour, $\mathcal C$, is taken around $t=0$.
Here $\mathcal R$ is either $S_m$ or $A_m$ and the
top sign refers to the symmetric case while the bottom sign refers to the antisymmetric case. 

We will focus just on evaluating $I^{S_m,A_m}$ now. Similar to \cite{Hartnoll:2006is}, we switch from
the complex plane to the complex cylinder by making the change of variables
\begin{equation}
t \to e^{\pi \sqrt{2\lambda} z}.
\end{equation}
The cylinder has periodicity $z=z+2i/\sqrt{2\lambda}$.
Moreover, since we are interested in the infinite rank limit of the Wilson loops, we introduce the variable
\begin{equation}
f \equiv \frac{m}{N},
\end{equation}
where $m$ is the rank of the representation, and we hold $f$ fixed in the large-$N$ limit.

\begin{figure}
\centering
\includegraphics[width=.5\textwidth]{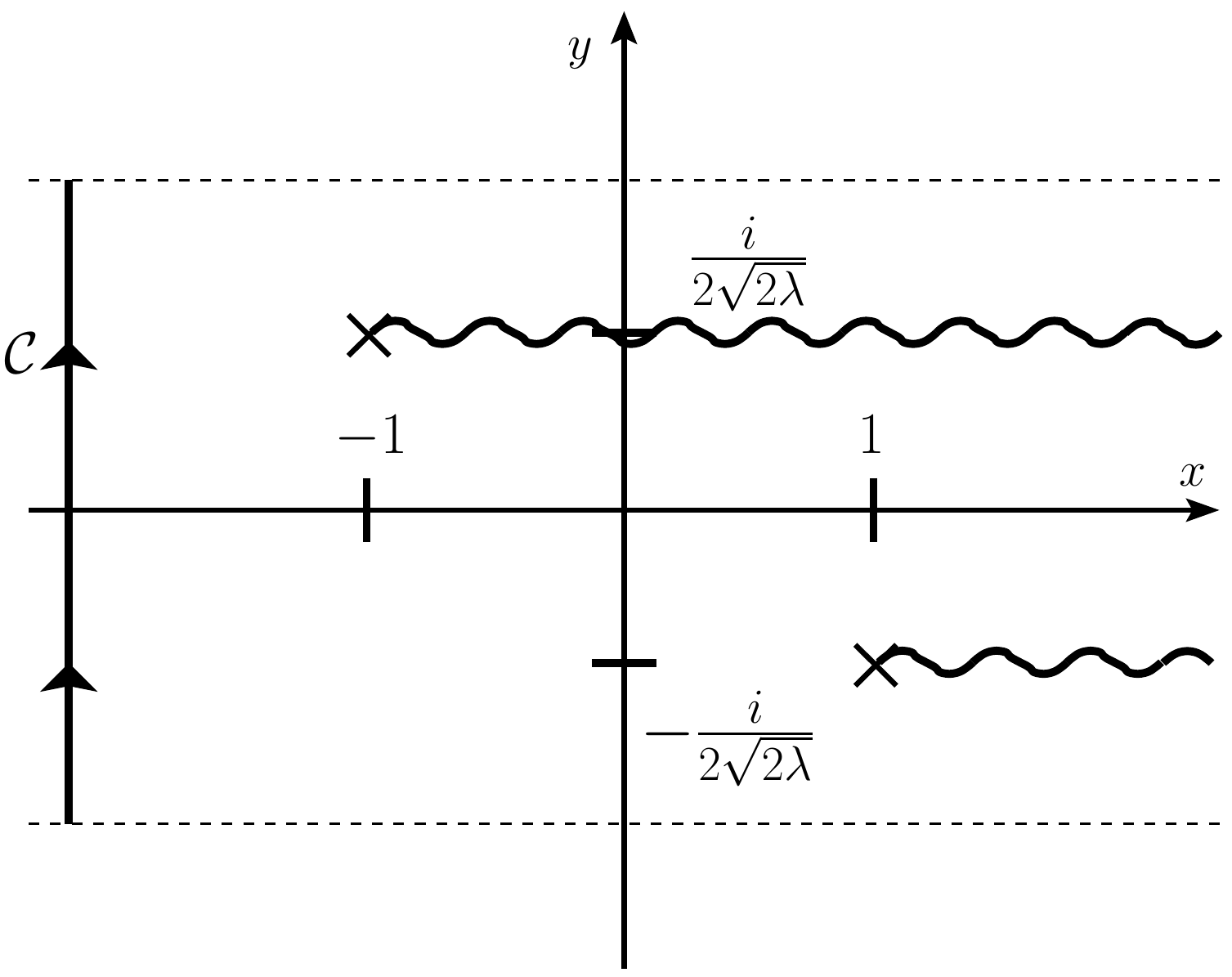}
\caption{We plot the branch cuts of the integrand of \eqref{Iint} in terms of $z=x+ i y$ for the antisymmetric case. The contour, $\mathcal C$, is shown to the left.}
\label{antisymcut}
\end{figure}

Because the eigenvalue density is uniform, the integral in the exponent can be explicitly carried out in terms
of dilogarithms yielding 
\begin{equation}
I^{S_m, A_m} = \frac{\sqrt{2\lambda}}{2 i} \oint_{\mathcal C}  dz \exp\left[\mp N\left(\frac{\Li_2\left(\mp ie^{\pi \sqrt{2\lambda}(z-1)}\right)-\Li_2\left(\pm ie^{\pi \sqrt{2\lambda}(z+1)}\right)}{2 \mu_* }\pm f \pi \sqrt{2\lambda} z
\right)\right].
\label{Iint}
\end{equation}
The principal branch of the dilogarithm, with
its branch cut along $[1,\infty)$, implies that the exponent has two branch cuts along
$z\pm \frac{i}{2\sqrt{2\lambda}} \in [-1,\infty)$ and 
$z\mp \frac{i}{2\sqrt{2\lambda}} \in [1,\infty)$, where the top sign is for the symmetric case as always (see Figure~\ref{antisymcut} and Figure~\ref{symcut}).
The contour, $\mathcal C$, lies to the left of the branch cuts.
We will now treat the two cases separately.

\section{Antisymmetric representation}\label{Sec:Am}

For the antisymmetric case, we take the bottom sign in \eqref{Iint}, and approximate the integral by steepest descent.  We find a saddle point at $\hat z$ where
\begin{equation}
e^{\pi\sqrt{2\lambda}\hat z} =\frac{\sinh (\mu_* f)}{\sinh(\mu_* ( 1-f))}.
\label{eq:aspzh}
\end{equation}
For large $\lambda$, and with $f \in (0,1)$, this expression becomes
\begin{equation}
\hat z \approx \fft{\mu_*}{\pi\sqrt{2\lambda}}(2f-1)=\left(1+ \frac{i}{2\sqrt{2\lambda}}\right)(2f-1).
\end{equation}
The saddle point is never near any of the branch points of the dilogarithms, so we may directly evaluate the  Gaussian integral at the saddle.

Expanding around the saddle point, we find that  the 
second derivative of the
integrand evaluated at the
saddle point is
\begin{align}
\frac{ \pi^2 \lambda}{\mu_*} \frac{\cosh (\pi \sqrt{2\lambda})}{\sinh (\pi \sqrt{2\lambda})-i\cosh(\pi \sqrt{2\lambda} \hat z)
}&=\frac{ \pi^2 \lambda}{\mu_*}\fft{\sinh(\pi\sqrt{2\lambda})+i\cosh((\pi\sqrt{2\lambda}+i\pi/2)(2f-1))}{\cosh(\pi\sqrt{2\lambda})}\nn\\
&\xrightarrow{\lambda \to \infty} \frac{\pi^2\lambda}{\mu_*}.
\end{align}
Inserting the saddle point value $\hat z$ into \eqref{Iint} and evaluating the Gaussian integral around the saddle then gives for the antisymmetric Wilson loop
\begin{equation}\label{Eq:Am}
\begin{aligned}
W_{1/6}^{A_m} &= \frac{-i}{\dim [A_m]}\sqrt{\frac{\mu_*}{N\pi}}\exp\left[
N \mu_* f(1-f)+\mathcal O(N/\mu_*)\right] \\
&\sim
\exp\left[
N \pi \sqrt{2\lambda} f(1-f)+\frac{1}{4}\ln\left( \frac{2\lambda}{N^2}\right)+\cdots\right].
\end{aligned}
\end{equation}
This result matches the calculation of the D6-brane in \cite{Drukker:2008zx} in the type IIA limit to leading order
 and has the expected  $f\to 1-f$ symmetry. We have also provided the first correction in the $1/N$ expansion to the saddle point result; it is the logarithmic term. Such term corresponds, in the holographic side, to a one-loop correction to the effective action of the dual D6 brane; partial results in this direction have recently been reported in \cite{Muck:2016hda}.

\section{Symmetric representation}\label{Sec:Sm}

\begin{figure}
\centering
\includegraphics[width=.5\textwidth]{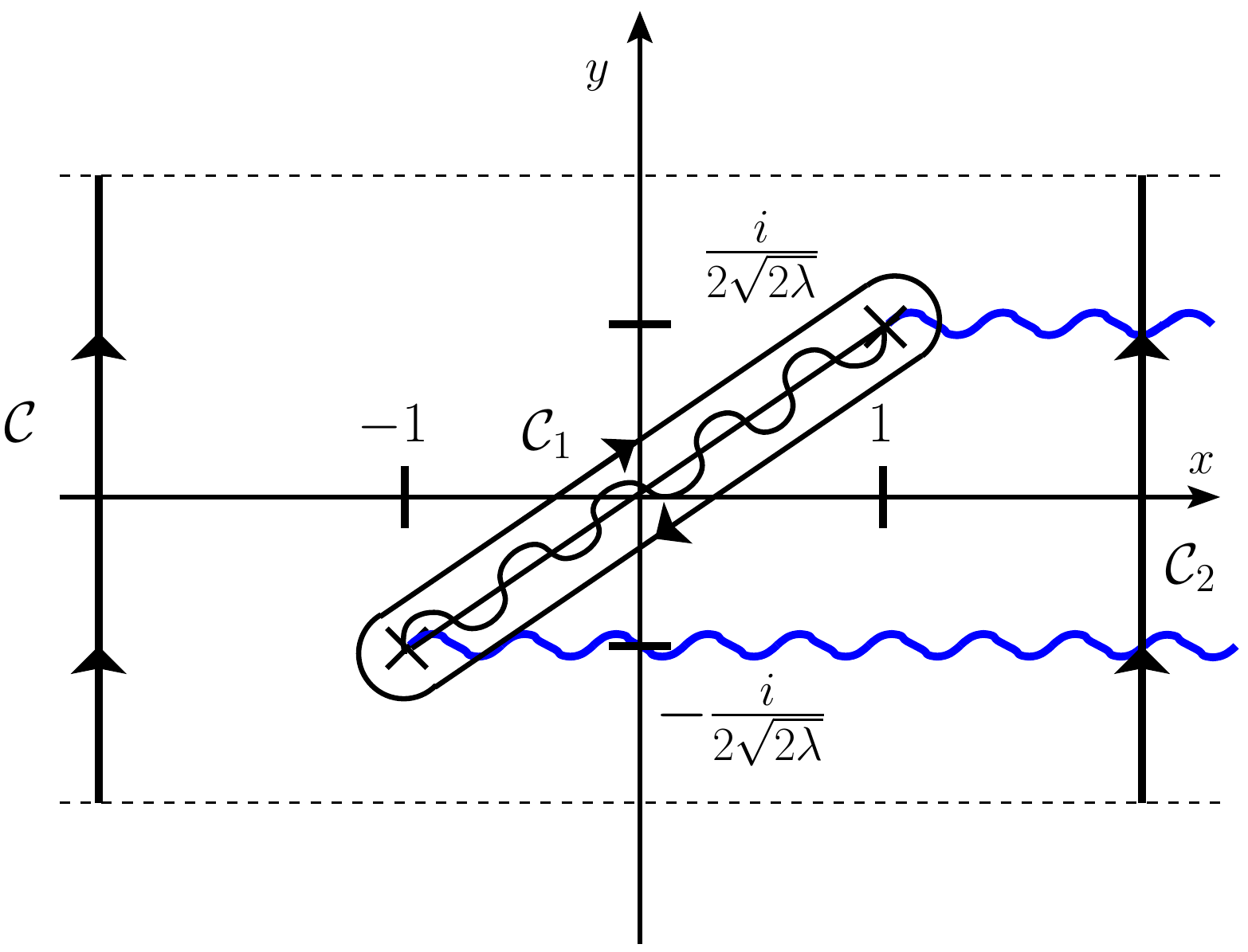}
\caption{The original branch cuts of \eqref{Iint} with $z=x+iy$ for the symmetric case are shown in blue, and the branch cut after the manipulation is shown in black. The integral reduces to just calculating the discontinuity across the branch cut, $\mathcal C_1$.}
\label{symcut}
\end{figure}

For the symmetric case, the steepest descent method leads to a saddle point exponentially close to the
branch point at $-1-i/2\sqrt{2\lambda}$, while the width of the Gaussian decreases slower than the
distance from the branch point.  Therefore, the saddle point approximation fails, and we must turn to
another method for evaluating the integral \eqref{Iint}.

In order to proceed, we find it convenient to deform the branch cut of the dilogarithm away from its
principal branch along $[1,\infty)$, and instead to lie along the curve $e^{\mu_* r}$ for $r \in [0,\infty)$.
As shown in Figure~\ref{symcut}, this allows the branch points of \eqref{Iint} to be joined by a single
branch cut that extends from $-1-{i}/{2\sqrt{2\lambda}}$ to $1+{i}/{2\sqrt{2\lambda}}$.  We may now
deform the original contour $\mathcal C$ into a new contour $\mathcal C'$ which is composed of two
parts: the discontinuity across the branch cut, $\mathcal C_1$, and a contour to the right of the branch cut,
$\mathcal C_2$. Because the integrand of \eqref{Iint} goes to 0 as $\Re[z] \to \infty$, the integral along
$\mathcal C_2$ vanishes. This allows us to write $I^{S_m}$ in terms of the discontinuity across the cut
encircled by $\mathcal C_1$:
\begin{equation}
I^{S_m} = \frac{\mu_* e^{N \mu_*f}}{2\pi i} \int_{0}^{2} dw \exp\left[-N\left(\frac{\Li_2\left(e^{\mu_*(w-2)}\right)-\Li_2\left(e^{\mu_*w}\right)}{2\mu_*}+ f \mu_* w\right)\right]
\left(1-e^{-N \pi i w}\right),
\label{wint}
\end{equation}
where the dilogarithms are taken on their principal branches.

In order to approximate the integral (\ref{wint}), we lift the factor $(1-e^{-N i \pi w})$ into the exponent and look for stationary points. The resulting saddle point equation is
\begin{equation}
\frac12\left[\ln(1-e^{\mu_*w})-\ln(1-e^{\mu_*(w-2)})\right] + f \mu_* + \frac{\pi i}{1-e^{N \pi i w}}=0,
\label{eq:spes}
\end{equation}
which is transcendental.  Numerically, we find a saddle point for $\mu_*w \sim 0$, so this leads us to assume
$|Nw|\ll1$, which implies $|\mu_*w|\ll1$ as well in the large-$N$ limit.  The first condition allows us to expand
the final term in (\ref{eq:spes}), while the second condition allows us to expand the logs.  
Dropping terms that are exponentially small and expanding for small $w$ then gives
\begin{equation}
\frac12 \ln (-\mu_* w) + \frac{i \pi}{2} + f \mu_* - \frac{\mu_*/N}{\mu_* w}=0.
\end{equation}
While this remains transcendental, it has the formal solution
\begin{equation}
\hat w=\fft{2/N}{W\left(\fft{2\mu_*}Ne^{2\mu_*f}\right)},
\label{eq:what}
\end{equation}
where the Lambert-$W$ function $W(z)$ is the inverse of $f(z)=ze^z$.  Unlike for the antisymmetric case, where
the saddle point expression (\ref{eq:aspzh}) depends only on $\mu_*$ and $f$, here there is also dependence on
$N$.  (This arises because of the last factor in (\ref{wint}) that encodes the discontinuity across the cut.)

In the IIA limit, where $\lambda$ and hence $\mu_*$ is held fixed, we may expand (\ref{eq:what}) in the large-$N$
limit to obtain
\begin{equation}
\hat w=\fft1{\mu_*}e^{-2\mu_*f}+\mathcal O(1/N),\qquad\mbox{(IIA limit)},
\end{equation}
where we have taken the principal branch of $W(z)$.  While this satisfies $|\mu_*\hat w|\ll1$,
the assumption that $|N\hat w|\ll1$ breaks down.  In this case, we would have to return to the full saddle
point expression (\ref{eq:spes}) in order to obtain $\hat w$.  Instead, we turn to the M-theory limit, where
$\hat\mu\sim\mathcal O(\sqrt{N})$, so that the argument of the Lambert-$W$ function becomes exponentially
large when $N\to\infty$.  In this case, we find
\begin{equation}
\hat w=\fft1{N\mu_*f}\left(1+\fft{\ln(Nf)-2\pi i}{2\mu_*f}+\cdots\right),\qquad\mbox{(M-theory limit)}.
\label{eq:Mtw}
\end{equation}
This is a self-consistent solution to the saddle point equation, as it satisfies both $|N\hat w|\ll1$ and
$|\mu_*\hat w|\ll1$.  We thus focus on the M-theory limit.

At this stage, we would ordinarily proceed with a saddle point approximation to the integral (\ref{wint}).
However, this problem has a moving maximum, $\hat w\sim 1/N$, which arises from the factor
$(1-e^{-N\pi i w})$ in (\ref{wint}).  Since the integral is dominated by the $w\to0$ limit, we instead expand
around $w=0$ using the relation
\begin{equation}
\Li_2(z) = - \Li_2(1-z)+\frac{\pi^2}6 - \ln(1-z) \ln z.
\end{equation}
The result is
\begin{equation}
I^{S_m} \approx \frac{\mu_* N}{2}
e^{N\mu_*f+\frac{N\pi^2}{12 \mu_*}}
 \int_{0}^{\infty} w\, dw \exp\left[-\fft{Nw}2\left(\ln\left(-\mu_* w\right)+2 \mu_*f -1\right)\right],
 \label{eq:expand}
\end{equation}
where we have discarded exponentially small terms in $\lambda$.  We have also extended the upper limit
of the integral to infinity, which only incurs an exponentially small error.  To proceed, we recall from
(\ref{eq:Mtw}) that the integrand is peaked at $\hat w\approx1/N\mu_*f$ in the complex plane.  We thus
make the substitution $w=z/N\mu_*f$, which results in the expression
%
\begin{equation}
I^{S_m} \approx \frac1{2 N\mu_* f^2}e^{N\mu_*f+\frac{N\pi^2}{12\mu_*}}
\int_{0}^{\infty} z\, dz \exp\left(-\alpha z-\fft{z\ln z}{2\mu_*f}\right),
\label{yint}
\end{equation}
where
\begin{equation}
\alpha=1-\fft{\ln(Nf)+i\pi+1}{2\mu_*f}.
\end{equation}
Note that we have deformed the contour in the complex plane in order to pass through the saddle
point that lies at a complex value of $\hat w$.

Because $|2\mu_*f|\gg 1$ as $N \to \infty$, the $z \ln z$ term in the exponent of (\ref{yint}) is
slowly varying. We therefore Taylor expand that part of the exponent to get:
\begin{equation}
I^{S_m}\approx \frac1{2 N\mu_* f^2}e^{N\mu_*f+\frac{N\pi^2}{12\mu_*}}
\int_{0}^{\infty} ze^{-\alpha z}\, dz \left(\sum_{n=0}^\infty
\frac{1}{n!}\left(-\frac{z \ln z}{2\mu_*f}\right)^n\right).
\end{equation}
Integrating one term of the sum at a time leads to an asymptotic expansion in $\mu_*\sim\sqrt{N}$. Evaluating
the first few terms gives
\begin{equation}
\begin{aligned}
I^{S_m}\approx \frac1{2 N\mu_* f^2}e^{N\mu_*f+\frac{N\pi^2}{12\mu_*}}\left[ \frac{1}{\alpha^2} - \frac{3 - 2 \gamma - 2 \ln\alpha}{2\mu_*f\alpha ^3}+\cdots\right],
\end{aligned}\label{jsmae}
\end{equation}
where $\gamma\approx 0.577$ is the Euler-Mascheroni constant. As $N \to \infty$, we see that the
expression in the square brackets on the right-hand side of \eqref{jsmae} approaches 1 (since $\alpha\to1$
in the limit).  As a result, the
expectation value of the Wilson loop in the symmetric representation takes the form
\begin{equation}\label{Eq:Sm}
\begin{aligned}
W_{1/6}^{S_m} & = \frac{1}{\dim[S_m]}\frac1{2 N\mu_*f^2}
\exp\left[N\mu_*f+\frac{N\pi^2}{12\mu_*}+o(1)\right]\\
& \sim \exp\left[N \pi \sqrt{2\lambda}f- \ln(2N \pi \sqrt{2\lambda}f^2)+\cdots\right],
\end{aligned}
\end{equation}
where in the second line we have kept only the leading terms in the large-$\lambda$ expansion.

At this point, it is worth recalling that the above expression was derived in the M-theory limit.  Working
in the IIA limit requires a different treatment, as the expression in (\ref{eq:expand}) was expanded
assuming $|N\hat w|\ll1$, which is violated in this limit.  Nevertheless, the leading behavior
$W_{1/6}^{S_m}\sim\exp(N\pi\sqrt{2\lambda}f)$ remains valid.  Furthermore, this
matches precisely the dual holographic calculation obtained as the classical action of a D2-brane embedded in $\textrm{AdS}_4\times \mathbb{CP}^3$ \cite{Drukker:2008zx}. To make the comparison with the holographic computations precise one needs to set $m=k/2$, where $k$ is the Chern-Simons level. As discussed in the old  literature of Chern-Simons theory,  the role of 't~Hooft operators affects the naive interpretation of the symmetric representation allowing values of $m$ only modulo $k$.  In the context of ABJM this effect remains and has been discussed, for example, in \cite{Hatsuda:2016rmv}.  Beyond the successful comparison with the leading result, it is worth noting that our answer contains, at subleading order,  a logarithmic term which would correspond to a one-loop computation on the holographic side. Some progress in the computation of the one-loop effective action of the dual D2-brane has recently been reported in \cite{Muck:2016hda}.

\section{Conclusions}\label{Sec:Conclusions}

In this manuscript we have computed the vacuum expectation value for $1/6$ supersymmetric Wilson loops  in the $m$-antisymmetric and the $m$-symmetric representations.  One important implication of our computation is that it matches the results obtained using the AdS/CFT correspondence. In particular, at leading order our matrix model computation equals the actions of the dual D6 and D2 brane configurations for the antisymmetric and symmetric representations, respectively, presented in \cite{Drukker:2008zx}. We have further computed sub-leading corrections which serve as a prediction for the holographic side. 

There has been  a concerted, decade-long, effort toward matching the holographic one-loop corrections with subleading terms in the field theory side  \cite{Forste:1999qn,Drukker:2000ep,Sakaguchi:2007ea,Kruczenski:2008zk,Kristjansen:2012nz}. More recently, in an attempt to tame  some of the intrinsic ambiguities on the holographic side, the ratio of 1/4 and 1/2 BPS Wilson loops has been compared to the field theory ratio \cite{Forini:2015bgo,Faraggi:2016ekd}, yielding some improvement in the comparison and pointing to interesting aspects of  string perturbation theory.  There is also an ongoing program of extending one-loop corrections to holographic configurations dual to Wilson loops in higher rank representations of the $SU(N)$  gauge group  in ${\cal N}=4$ SYM \cite{faraggi:2011bb,Faraggi:2011ge,Buchbinder:2014nia,Faraggi:2014tna}.  
Among the open problems that our work stimulates, a logical continuation will include comparing our results with  the one-loop effective actions of the dual  D2 and D6 configurations. Some partial results toward the one-loop effective action of the corresponding dual D6 and D2 branes have recently been reported in \cite{Muck:2016hda}. The work presented in this manuscript can be viewed as an important step into extending  similar high precision comparisons to the context of the $\textrm{AdS}_4\times \mathbb{CP}^3$/ABJM correspondence. 

Another outstanding problem to which we hope to return pertains to the various order of limits presented here. One lesson that can be drawn from \cite{Faraggi:2014tna} is that in attempting  to match with the holographic results it is important to understand clearly various orders of limits. Some progress in this direction has recently been reported in the context of the ${\cal N}=4$ SYM Gaussian matrix model \cite{Horikoshi:2016hds}. In this manuscript we have largely restricted ourselves to the M-theory limit and have verified that its limit of applicability goes beyond what is reasonable  to expect in the sense that we are able to match, at leading order,  holographic results for the D6 and D2 branes corresponding to the IIA limit. Along the same lines, It might also be interesting to use advanced matrix model techniques with the aim of finding exact solutions for any set of $(N, k, m)$.

\section*{Acknowledgments}
We thank Diego Trancanelli for various comments and discussions.
This material is based upon work supported by the National Science Foundation under Grant No.~PHY~1559988 and by the US Department of Energy under Grant No.~DE-SC0007859.

\appendix

\section{A note on the matrix eigenvalue distribution on the ABJM slice}
\label{App:Numerical}

It is worth emphasizing that the eigenvalue distribution (\ref{eq:rhomu}) was initially derived for the lens space matrix model in the
planar limit.  In particular, the eigenvalues $\mu_i$ and $\nu_i$ are taken to be real, and condense along
two cuts in the complex plane.  (When working with the total resolvent, the $\nu_i$ cut is displaced by $i\pi$,
although the eigenvalues themselves remain real.)  The analytic continuation to the ABJM slice is then
accomplished by taking the `t~Hooft parameters to be imaginary
\begin{equation}
t_1=-t_2=2\pi i\lambda.
\end{equation}
The resulting expression for the density $\rho(\mu)$ is then
\begin{equation}
\rho(z)=\fft{-i}{2\pi^2\lambda}\tan^{-1}\left(\sqrt{\fft{2+i\kappa-2\cosh z}{2-i\kappa+2\cosh z}}\right),
\label{eq:rhoz}
\end{equation}
where
\begin{equation}
\lambda=\fft\kappa{8\pi}{}_3F_2(\ft12,\ft12,\ft12;1,\ft32;-\kappa^2/16).
\end{equation}
The range of $z$ in (\ref{eq:rhoz}) is taken between $-\mu_*$ and $\mu_*$ where
\begin{equation}
\mu_*=\ln\left[\fft12\left(2+i\kappa+\sqrt{\kappa(4i-\kappa)}\right)\right].
\label{eq:zstar}
\end{equation}
The points $-\mu_*$ and $\mu_*$ are branch points of the ABJM resolvent.  However, the location of the cut is
not entirely obvious when continued away from the real axis.  Since $\rho(z)$ is analytic away from the branch
points, saddle point integrals, such as (\ref{eq:FAFS}), remain valid when continued from the lens space matrix
model to the ABJM slice.  Nevertheless, it is instructive to examine the eigenvalue distribution in the
complex plane.

We may gain insight on the $\mu_i$ and $\nu_i$ distribution in the planar limit of ABJM theory by
numerically solving the saddle point equations arising from the partition function (\ref{eq:ABJMZ})
\begin{align}
\fft{-ik}{2\pi}\mu_i&=\sum_{j\ne i}\coth\fft{\mu_i-\mu_j}2-\sum_j\tanh\fft{\mu_i-\nu_j}2,\nonumber\\
\fft{ik}{2\pi}\nu_i&=\sum_{j\ne i}\coth\fft{\nu_i-\nu_j}2-\sum_j\tanh\fft{\nu_i-\mu_j}2.
\label{eq:Forces}
\end{align}
This was investigated in the M-theory limit in \cite{Herzog:2010hf} by treating the above equations
as effective forces acting on the eigenvalues.  Alternatively, it is possible to perform multi-dimensional
root finding directly within modern computer algebra systems (such as using FindRoot[] in
Mathematica).

\begin{figure}[t]
\begin{center}
\includegraphics[width=7cm]{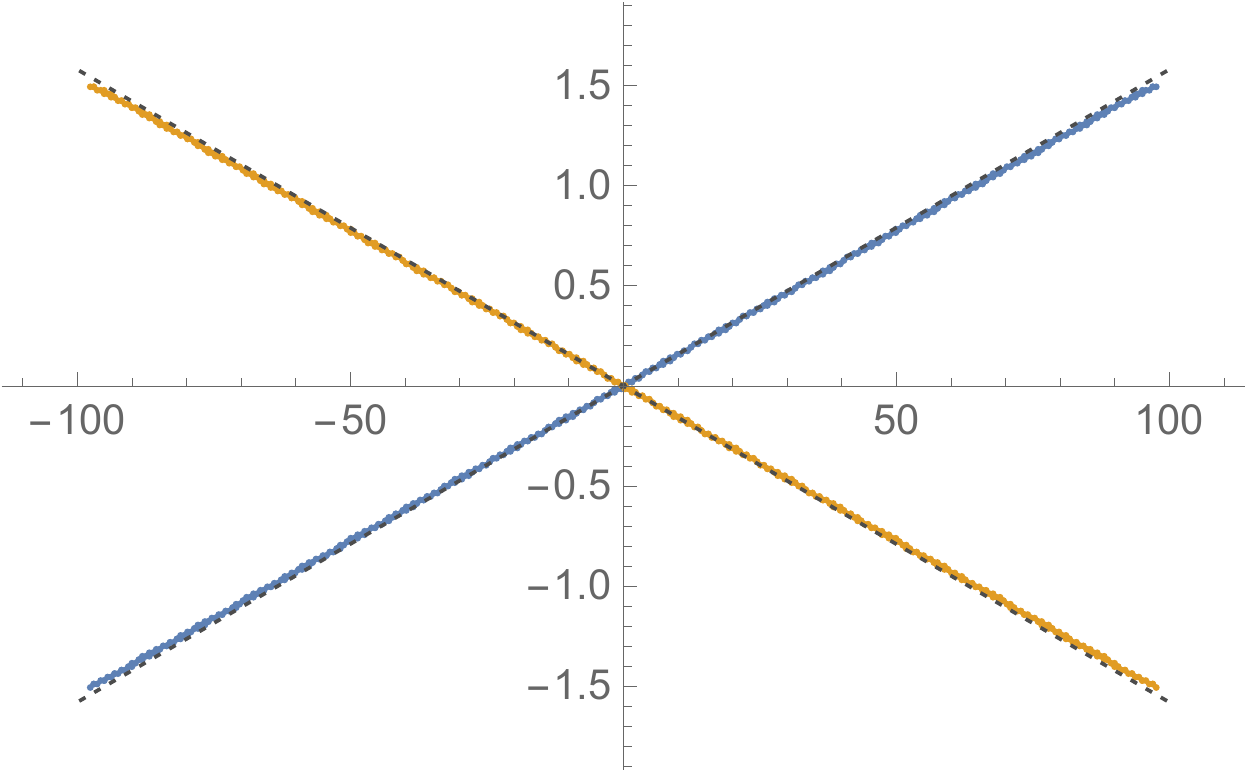}
\kern.5cm
\includegraphics[width=7cm]{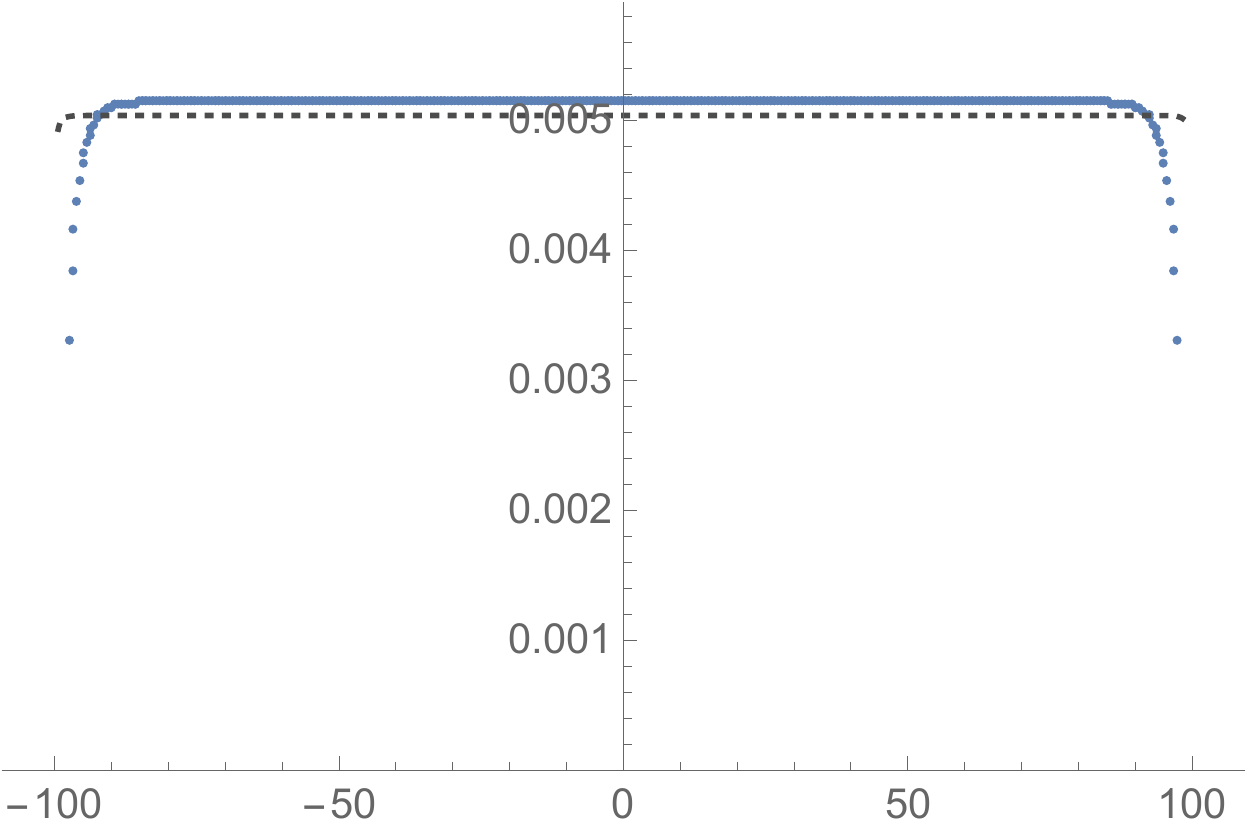}
\end{center}
\caption{The numerical eigenvalue distribution for $N=500$ and $k=1$.  Left: distribution of $\mu_i$ and
$\nu_i$ in the complex plane.  Right: the density $\hat\rho(x)$ of the real part of the $\mu_i$ eigenvalues.
\label{fig:k=1}}
\end{figure}

If we ignore the coupling between the $\mu_i$ and $\nu_i$, then the equilibrium position of the
eigenvalues is governed by the force balance between a harmonic oscillator potential (albeit with
imaginary spring constant $\pm k/2\pi i$) and a $\coth$ repulsion between eigenvalue pairs.  In the
M-theory limit, the repulsion dominates over the harmonic oscillator, and the eigenvalues are spread
out.  In this case, the $\coth$ in (\ref{eq:Forces}) can be approximated by $\pm1$ depending on the
relative ordering of the eigenvalues.  Balancing this against a linear Hooke's law force then gives
rise to a uniform distribution \cite{Herzog:2010hf} along the line connecting $-\mu_*$ to $\mu_*$
where $\mu_*=\pi\sqrt{2\lambda}+i\pi/2$ for the $\mu_i$ eigenvalues, and the complex conjugate
for the $\nu_i$ eigenvalues.  The comparison with the numerical solution is shown in Fig.~\ref{fig:k=1}.

\begin{figure}[t]
\begin{center}
\includegraphics[width=7cm]{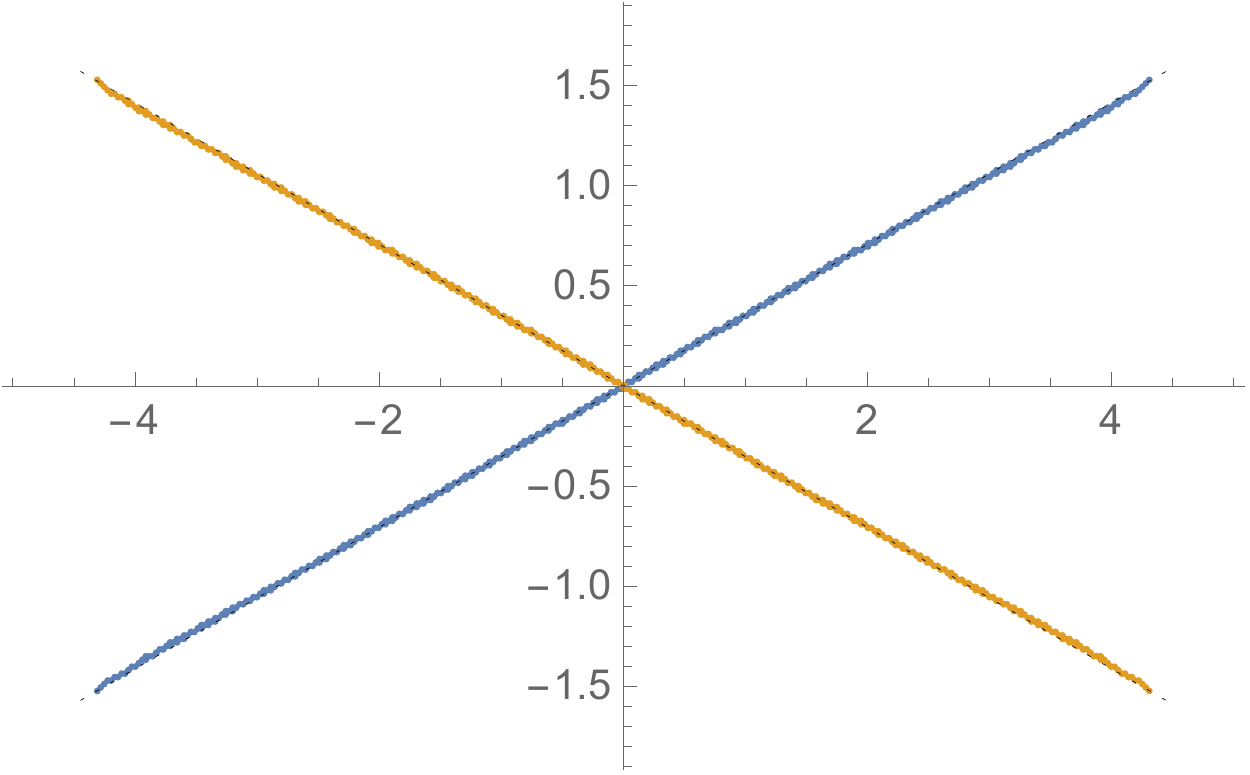}
\kern.5cm
\includegraphics[width=7cm]{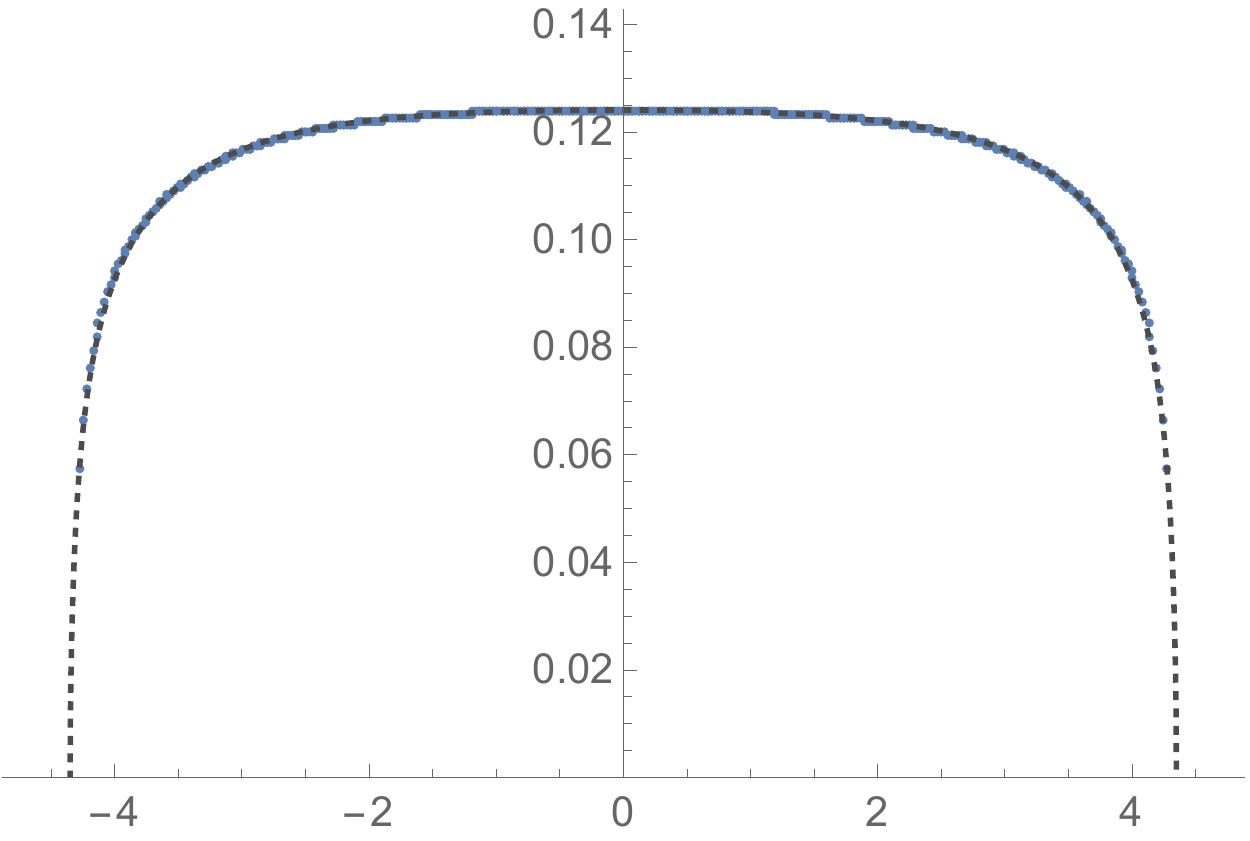}
\end{center}
\caption{The numerical eigenvalue distribution for $N=500$ and $\lambda=1$.  Left: distribution of
$\mu_i$ and $\nu_i$ in the complex plane.  The dashed line corresponds to the linear approximation
$\mu_i\in[-\mu_*,\mu_*]$ and its complex conjugate for $\nu_i$ where $\mu_*$ is given in (\ref{eq:zstar}).
Right: the density $\hat\rho(x)$ of the real part of the $\mu_i$ eigenvalues.  The dashed line 
corresponds to the real part of $\hat\rho$ given in (\ref{eq:rhohat}).
\label{fig:k=500}}
\end{figure}

We are, of course, mostly interested in the IIA limit.  In this case, as $k$ is increased, the harmonic oscillator
spring constant gets stronger, and the eigenvalues are pulled closer towards the origin.  As a result,
the approximation $\coth\sim\pm1$ is no longer valid, and as $\lambda\to0$ the $\coth$ repulsion in
(\ref{eq:Forces}) can be better approximated as $\coth z\sim 1/z$.  Since this matches the repulsion
arising from the Vandermonde determinant in the ordinary Hermitian matrix model, the eigenvalue
distribution more closely resembles the Wigner semi-circle distribution in the small $\lambda$ limit,
at least when suitably translated into the complex plane.  As an example, we show the numerical
solution for $N=500$ and $\lambda=1$ in Fig.~\ref{fig:k=500}.

\begin{figure}[t]
\begin{center}
\includegraphics[width=7cm]{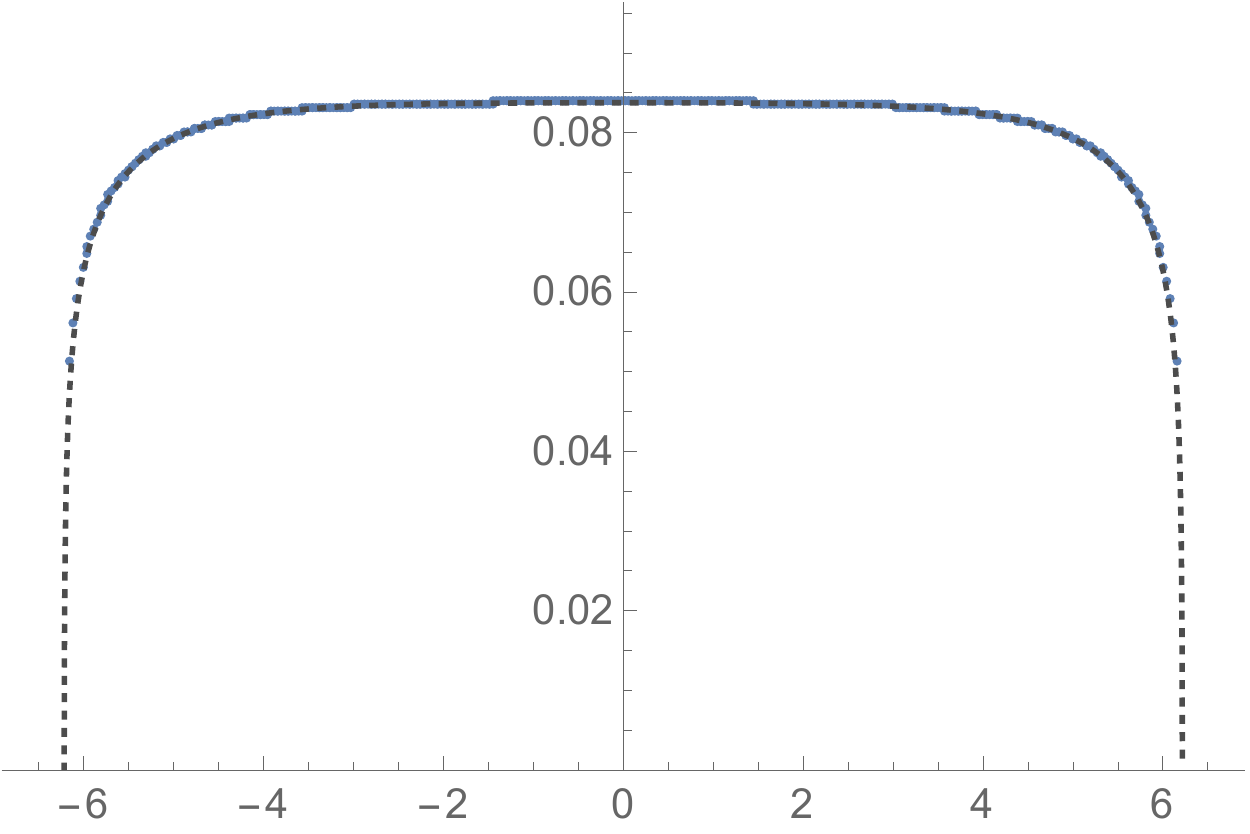}
\kern.5cm
\includegraphics[width=7cm]{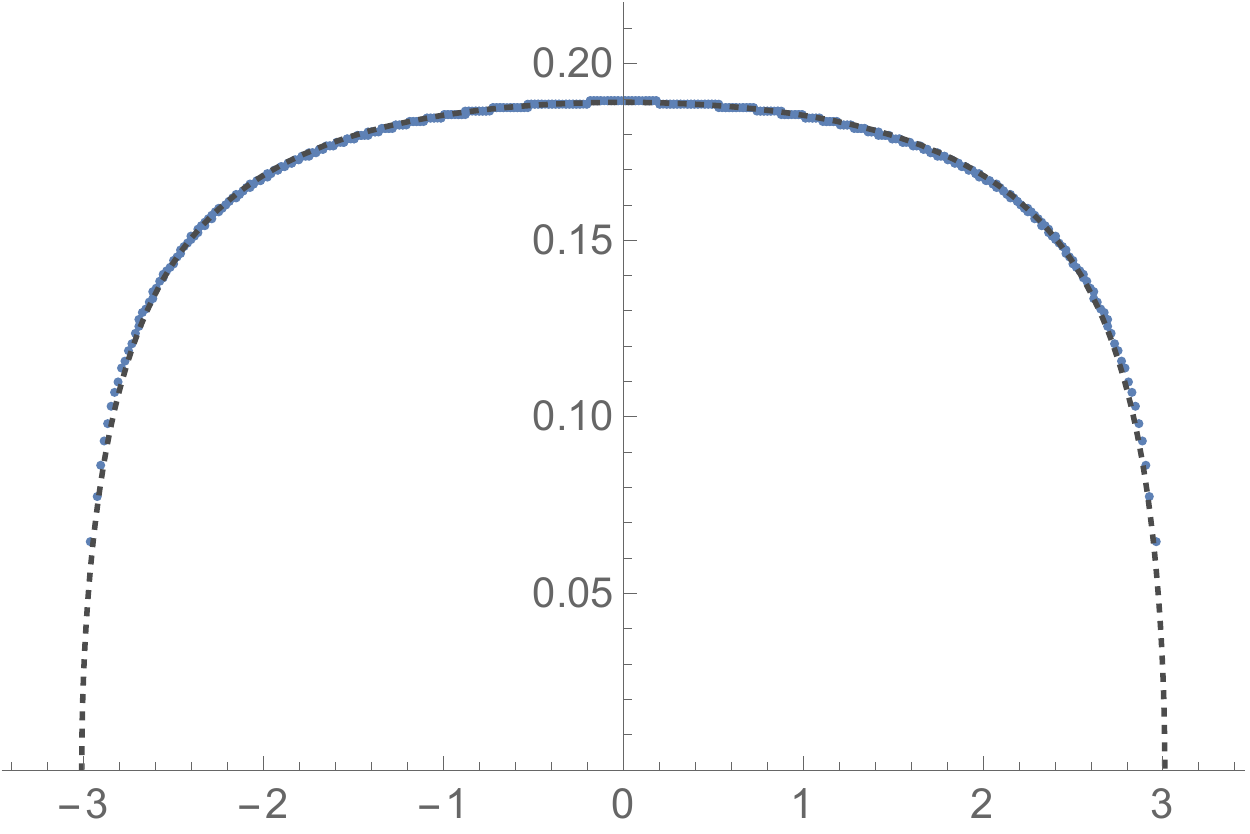}
\end{center}
\caption{The density $\hat\rho(x)$ of the real part of the $\mu_i$ eigenvalues for $N=500$.  The left
plot corresponds to $\lambda=2$ and the right plot corresponds to $\lambda=1/2$.  (The
$\lambda=1$ plot is shown in Fig.~\ref{fig:k=500}.)  The dashed lines corresponds to the real part
of $\hat\rho$ in (\ref{eq:rhohat}).
\label{fig:k=250,1000}}
\end{figure}

In order to obtain a real density of eigenvalues, note that the eigenvalue density $\rho(\mu)$ in
(\ref{eq:rhoz}) is normalized according to
\begin{equation}
\int_{-\mu_*}^{\mu_*}\rho(z)dz=1,
\end{equation}
where the integral is taken along the cut where the eigenvalues lie.  Writing $z=x+iy$, we may convert
this to a real integral
\begin{equation}
\int_{-x_*}^{x_*}\rho(x+iy(x))(1+i y'(x))dx=1.
\end{equation}
The path $y(x)$ describes the cut, and can in principle be solved for by demanding that the above
integrand, which represents the density of the real part of the eigenvalues, be real along the cut.
We have not actually solved for the actual path.  However, examination of Fig.~\ref{fig:k=500} shows
that the cut remains essentially a straight line segment joining $-\mu_*$ to $\mu_*$ (for the $\mu_i$
distribution).  We may thus take $y(x)=({y_*}/{x_*})x$, so that
\begin{equation}
\hat\rho(x)\equiv \rho(x+iy(x))(1+i y'(x))=\zeta\rho(\zeta x),\qquad
\zeta=\fft{z_*}{\Re[z_*]}=1+i\fft{y_*}{x_*}.
\label{eq:rhohat}
\end{equation}
Note that $\hat\rho$ generally has a small imaginary component, indicating that the straight line
cut approximation is not exact.  Nevertheless, $\Im[\hat\rho]$ is suppressed in the large $\lambda$ limit,
and $\Re[\hat\rho]$ agrees well with the numerical results.  Additional eigenvalue densities for
$\lambda=2$ and $\lambda=1/2$ are shown in Fig.~\ref{fig:k=250,1000}.

\bibliographystyle{JHEP}
\bibliography{WLoops-bib}

\end{document}